\documentclass[10pt, conference, compsocconf]{IEEEtran}

\usepackage{cite}

\ifCLASSINFOpdf
\usepackage[pdftex]{graphicx}
\else
\fi
\usepackage[cmex10]{amsmath}
\usepackage{algorithmic}

\usepackage[font=footnotesize]{caption}
\usepackage{url}

\usepackage{booktabs}
\usepackage{adjustbox}
\usepackage{color}
\usepackage{algorithm}
\usepackage{subcaption}
\usepackage{multirow}
\let\labelindent\relax
\usepackage{enumitem}
\usepackage{setspace}
\usepackage{amsfonts}

\let\OLDthebibliography\thebibliography
\renewcommand\thebibliography[1]{
  \OLDthebibliography{#1}
  \setlength{\parskip}{0pt}
  \setlength{\itemsep}{0pt }
}

\begin{document}

\title{In-Depth Exploration of Single-Snapshot Lossy Compression Techniques for N-Body Simulations\vspace{-1mm}}

\author{\IEEEauthorblockN{Dingwen Tao,\IEEEauthorrefmark{1}
Sheng Di,\IEEEauthorrefmark{2}
Zizhong Chen,\IEEEauthorrefmark{1}\IEEEauthorrefmark{3} and
Franck Cappello\IEEEauthorrefmark{2}\IEEEauthorrefmark{4}}
\IEEEauthorblockA{\IEEEauthorrefmark{1}
University of California,
Riverside, CA, USA\\ \{dtao001, chen\}@cs.ucr.edu}
\IEEEauthorblockA{\IEEEauthorrefmark{2}Argonne National Laboratory, IL, USA\\
\{sdi1, cappello\}@anl.gov}
\IEEEauthorblockA{\IEEEauthorrefmark{3}Beijing University of Technology, Beijing, China}
\IEEEauthorblockA{\IEEEauthorrefmark{4}University of Illinois at Urbana-Champaign, IL, USA}
 \\[-2.0ex]
}


%


\maketitle

\begin{abstract}
\linespread{0.92}\selectfont
In situ lossy compression allowing user-controlled data loss can significantly reduce the I/O burden.
For large-scale N-body simulations where only one snapshot can be compressed at a time, the lossy compression ratio is very limited because of the fairly low spatial coherence of the particle data. 
In this work, 
we assess the state-of-the-art single-snapshot lossy compression techniques of two common N-body simulation models: cosmology and molecular dynamics.
We design a series of novel optimization techniques based on the two representative real-world N-body simulation codes.
For molecular dynamics simulation, we propose three compression modes (i.e., best\_speed, best\_tradeoff, best\_compression mode) that can refine the tradeoff between the compression rate (a.k.a., speed/throughput) and ratio. 
For cosmology simulation, we identify that our improved SZ is the best lossy compressor with respect to both compression ratio and rate.
Its compression ratio is higher than the second-best compressor by 11\% with comparable compression rate.
Experiments with up to 1024 cores on the Blues supercomputer at Argonne show that our proposed lossy compression method can reduce I/O time by 80\% compared with writing data directly to a parallel file system and outperforms the second-best solution by 60\%.
Moreover, our proposed lossy compression methods have the best rate-distortion with reasonable compression errors on the tested N-body simulation data compared with state-of-the-art compressors.
\end{abstract}

%
\IEEEpeerreviewmaketitle

\section{Introduction}

Because of ever-increasing parallel execution scale, today's scientific simulations are producing volumes of data too  large to be accommodated in storage systems. The limitation comes from the limited storage capacity and I/O bandwidth of parallel file systems in production facilities. Cosmology simulations such as the Hardware/Hybrid Accelerated Cosmology Code (HACC) \cite{habib2016hacc} are typical examples of parallel executions facing this issue. HACC solves an N-body problem involving domain decomposition, a medium-/long-range force solver based on a particle-mesh method, and a short-range force solver based on a particle-particle/particle-mesh algorithm. According to cosmology researchers, the number of particles to simulate can be up to 3.5 trillion in today's simulations (and even more in the future), which leads to 60 PB of data to store; yet a system such as the Mira  supercomputer has only 26 PB of file system storage. Currently, HACC users rely on decimation in time, storing only a fraction of the simulation snapshots, to reduce the pressure on the storage system. A reduction factor of 80\% to 90\% is commonly used. At exascale, temporal decimation will not be enough to address the limitations of the storage system: snapshots will be so large (each in the range of 5 PB) that the time to store each snapshot (83 minutes on a storage system offering a sustained bandwidth of 1 TB/s) will become a serious problem. HACC is not a special case. Another example of exascale particle simulation requiring data reduction is the atomistic modeling of the microstructural evolution of material. This application uses the Accelerated Molecular Dynamics Family (AMDF) code \cite{amcf} developed at Los Alamos National Laboratory for generating high-quality trajectories of ensembles of atoms in materials. At exascale, storing each full snapshot in this case would also take too long, however, so that in situ compression of each snapshot is needed. 

In this paper, we focus on how to optimize the lossy compression quality of individual snapshots for scientific data sets produced by N-body simulations. Compressing individual snapshots is challenging because we cannot reduce the data size by leveraging the smoothness of particle's trajectory (such as  compression in the time dimension), which is the common  approach used to  compress particle data sets. Instead, we can use only the spatial information and fairly limited correlation of adjacent elements in the data set to perform the data compression. This approach is significantly more challenging than mesh-data compression and trajectory-based data compression. Unlike the regular or irregular mesh data generated by conventional simulations such as fluid dynamics, the data fields of particles such as the coordinate and velocity are stored in separate 1D arrays across different dimensions. In the two applications considered in this study, the indices of every 1D array are kept consistent for the same particle. All key data in the HACC code, for example, are stored in six 1D arrays:  three coordinate fields (xx, yy, zz) and three velocity fields (vx, vy, vz). Because of the lack of correlation between adjacent particles in the data set, state-of-the-art lossy compressors (such as SZ \cite{sz16, sz17} and ZFP \cite{zfp}) cannot work effectively. Moreover, the data across various fields are  diverse, with different features.

In this work, 
we assess the existing state-of-the-art single-snapshot lossy compression techniques based on the two common N-body simulation models: the AMDF molecular dynamics simulation model and HACC hierarchical cosmology simulation model.
The AMDF code is a classical molecular dynamics code developed by LANL, and HACC is a typical representative of the hierarchical model.
We propose a series of novel optimization techniques. 
Specifically, we make the following four contributions.
\begin{itemize}[noitemsep,topsep=4pt]
\item We propose three compression modes for molecular dynamics simulation that can refine the tradeoff between the compression rate (a.k.a., speed/throughput) and ratio. 
(1) Our solution with best\_speed mode (called SZ-LV) improves the prediction method, such that the compression rate turns 4.4x as fast as that of the existing best-ratio compressor CPC2000 with only 12\% lower compression ratio. (2) The best\_tradeoff mode (called SZ-LV-PRX) adopts a partial-radix sorting based on R-index, obtaining 2x improvement on the compression rate with unchanged compression ratio compared with CPC2000. (3) The best\_compression mode (called SZ-CPC2000) combines our improved SZ and CPC2000, increasing the compression ratio by 13\% and the compression rate by 10\% over CPC2000.
\item For cosmology simulation, we identify that our SZ-LV is the best lossy compressor with respect to both compression ratio and rate. Its compression ratio is higher than the second-best compressor by 11\% with comparable compression rate.
\item Our experiments with up to 1024 cores on a supercomputer show that our proposed lossy compression method can reduce I/O time by 80\% compared with writing data directly to a parallel file system and outperforms the second-best solution by 60\%. The experiments also show that our proposed lossy compression methods have the best rate-distortion with reasonable compression errors on the tested N-body simulation data compared with state-of-the-art compressors.
\end{itemize}

The rest of the paper is organized as follows. We discuss the related work in Section \ref{sec: related}. In Section \ref{sec: problem}, we formulate the data compression problem based on the N-body simulation data sets. In Section \ref{sec: assess}, we assess the existing state-of-the-art lossy compression approaches (or compressors) on the two real-world N-body simulation data. 
In Section \ref{sec: optimization}, we propose our novel optimization methods to improve the lossy compression for the N-body simulation data. In Section \ref{sec: evaluation}, we evaluate the compression ratios and compression rates for all the lossy compression approaches based on both serial execution and parallel execution. We provide concluding remarks in Section \ref{sec: conclusion}.

\section{Related Work}
\label{sec: related}

\paragraph{Lossless and Lossy Compression}
Data compression has been extensively studied for decades and can be split into two categories: lossless compression and lossy compression. Lossless compressors (such as Huffman coding \cite{huffman} and LZ77 \cite{lz77}) can work effectively on the digital streams (such as image or sound) that are composed of integer numbers. However, they suffer from a very low compression ratio on scientific data sets composed of floating-point values (as confirmed by \cite{sz16,sz17}), because of high randomness of the tailing mantissa bits such that finding the exact same patterns or repeated values in the sequence is difficult. 

Recently, many lossy compressors \cite{numarck,ssem,isabela,fpzip,zfp,sz16,pattern} have been designed and implemented for scientific data. Many of them are designed for mesh data sets, which are expected to have strong coherence among the nearby data in the data set. SZ \cite{sz16}, for example, predicts the value for each data point by its preceding neighbors in the multidimensional space and then performs a customized Huffman coding \cite{huffman} to shrink the data size significantly. ZFP \cite{zfp} splits the whole data set into many small blocks with an edge size of 4 along each dimension and compresses the data in each block separately by a series of carefully designed steps (including alignment of exponent, orthogonal transform, fixed-point integer conversion, and binary representation analysis with bit-plane encoding). FPZIP \cite{fpzip} adopts predictive coding and also ignores insignificant bit planes in the mantissa based on the analysis of IEEE 754 binary representation \cite{ieee2008754}. SSEM \cite{ssem} splits data into a high-frequency part and low-frequency part by wavelet transform \cite{wavelet} and then uses vector quantization and GZIP. ISABELA \cite{isabela} sorts the data and then performs the data compression by B-spline interpolation. Several papers \cite{Petereval, ssem} have shown that these lossy compressors can work effectively on mesh data, but the quality of their compression declines on N-body simulation data sets, which is a significant gap for the compressors and N-body simulation researchers.

\paragraph{N-body Simulation Data Compression}
Compression of N-body simulation data sets has also been studied for years, but most of the methods proposed are designed based on smooth temporal trajectory of the same particles \cite{Yang-sc1999, ed2006, Kumar2013, numarck}.
All such trajectory-based compression schemes requires to load/keep multiple snapshots during the compression/simulation, hence, they are not suitable for extremely large-scale simulation in which only one snapshot is allowed to be loaded into the memory.

In addition to the temporal-coherence based compression methods, we noted one related work on the particle compression based on single snapshot.  Omeltchenko et al. \cite{cpc2000} proposed a lossy compression method (called CPC2000 in this paper) that does not rely on the temporal coherence. Four steps are involved: (1) It first converts all floating-point values (including coordinate fields and velocity fields) to integer numbers by dividing them by user-required error bound. (2) Then, it reorganizes all particles in the space onto a zigzag-similar space-filling curve \cite {zigzag} by splitting the space into multiple small blocks and constructing a uniform Oct-tree index in each block based on the coordinate fields $(xx,yy,zz)$. (3) It sorts the particles based on the R-indices by a radix-similar sorting method in each block, and then computes the difference of the adjacent indices. (4) It performs a tailored adaptive variable-length encoding to shrink the storage size. We implemented the method rigorously and compare it with other approaches in our evaluation section.

\section{Problem Statement}
\label{sec: problem}

As discussed above, the main limitation of lossless compressors is their limited data reduction capability \cite{ratana} and even lower on N-body simulation data sets, thus we focus on the lossy compression methods for N-body simulations. As discussed above, because of considerations of memory consumption, we focus on single-snapshot compression without utilization of temporal coherence in this work. 

Compression in parallel falls into two categories: in situ compression and off-line compression. For in situ compression, each rank or process can compress/decompress a fraction of the data that is being held in its memory. For off-line compression, a parallel program can be used to load the data into multiple processes and run the compression separately on them. In this work, we focus on the in situ compression for parallel N-body simulation application rather than off-line compression.
 
N-body simulations contain many variables each representing one data field of particles. In the two applications considered in this study, the variables are stored in separate 1D arrays. Each array has a specific data type, for example, integer, single or double floating point, and string data. Since the major type of the scientific data is floating point, we focus on floating-point data compression. Note that the variables may be stored in an array of structures or a structure of arrays in other N-body simulations, but the storage format doe not change the data compression problem. 
Further, each snapshot of N-body simulations contains six floating-point variables, namely, $xx, yy, zz, vx, vy, vz$. The first three indicate coordinate information, and the other three indicate velocity along the three dimensions. As confirmed by HACC and AMDF code developers/users, the only floating-point data used in the analysis by them are positions and velocities and other quantities of interest can be recomputed from positions and velocities. Also, other N-body simulations may use other fields such as energy and entropy that are much smoother and easier to compress than position and velocity, such that the final size of the compressed data set is largely dominated by position and velocity compressibility. Therefore, we focus on optimizing the lossy compression of \textit{coordinate and velocity data} in N-body simulation.

In summary, the main objective of our work is to \textit{optimize the single-snapshot in situ lossy compression quality for N-body simulation data sets, mainly on coordinate and velocity data}, provided that the compression errors are controlled within a user-specified bound for each data point. \textit{Compression ratio} (i.e., compression factor) is a key assessment metric; it is the ratio of the original data size to the compressed data size. The distortion of the data through the compression can be evaluated by different measurements, such as \textit{maximum point-wise compression error} and \textit{peak-to-signal ratio}. Moreover, an ideal in situ data compressor should also have high \textit{compression rate}.

To describe the evaluation metrics clearly, we use the particle's velocity variable $vx$ as an example and define some necessary notation as follows. We denote the original 1D array of $vx$ as $\{vx_1,vx_2,..., vx_N\}$, where each $vx_i$ is a floating-point scalar. We denote the reconstructed array  by $\{\tilde{vx_1},\tilde{vx_2},..., \tilde{vx_N}\}$, which is recovered by the decompression process. We also denote the value range of variable $vx$ by $R_{vx}$, that is, $R_{vx} = vx_{max} - vx_{min}$. 

\paragraph{Pointwise compression error} For data point $i$, let $e_{i} = |vx_i - \tilde{vx_i}|$, where $e_{i}$ is the \textit{absolute error}. For error-controlled lossy compression, the compression errors can be guaranteed within an error bound, which can be expressed as $e_{i} < eb_{abs}$ for $1\leq i \leq N$, where $eb_{abs}$ is a user-specified absolute error bound. Similar to \cite{sz17}, we also define the \textit{value-range-based relative error bound} as the ratio of $eb_{abs}/R_{vx}$, denoted by $eb_{rel}$.

\paragraph{Average compression error} To evaluate the average error in the compression, we adopt the normalized root mean squared error (NRMSE), which is calculated as 
$\sqrt{\sum_{i=1}^{N} e_{i}^2 / N}/R_{vx}$.

\paragraph{Compression and decompression rate} To evaluate the speed/throughput of the compressor, we calculate the compression and decompression rate (in bytes/second) based on execution time and data size.

\section{Overview of State-of-the-art Compression Methods on N-body Simulation Data Sets}
\label{sec: assess}

Before proposing our analysis and optimizations, we first evaluate the compression ratios of the current state-of-the art lossless and lossy compressors including GZIP, CPC2000, FPZIP, ISABELA, ZFP, and SZ. 
Since SSEM cannot guarantee compression errors within user-set error bounds \cite{sz16} and since NUMARCK depends on the multiple snapshots \cite{numarck}, they are not included in this research and we do not present their compression ratios here.
We conduct the compression on two real-word N-body simulation data sets: \textit{HACC} data sets from cosmology simulation and \textit{AMDF} data sets from shape evolution simulation of small platinum nanoparticles. Both of the two data sets are single precision floating-point data. The details of the applications are shown in Table \ref{tab: apps}. Note that, although FPZIP, ISABELA, ZFP, and SZ are designed for mesh data, we directly perform their compression on separate 1D arrays storing data with consistent indices for particles, because it is the unique way to handle them by existing mesh-based compressors.
For the lossless compressor GZIP, we use the best-ratio mode. For the lossy compressors, because of space limitations, we show only the typical error bound, that is, the value-range-based relative error bound $10^{-4}$, which is accurate enough for data analysis as stated by users. Specifically, for CPC2000, ZFP, and SZ, we use the absolute error bound computed based on $eb_{rel} = 10^{-4}$ and the data value range for each variable. Note that ZFP has three compression modes including fixed bit-rate, fixed precision, and fixed accuracy. As suggested by ZFP's developer, we adopt the best mode (i.e., fixed accuracy) with respect to compression ratio. For ISABELA, we use the pointwise relative error bound computed based on $eb_{rel} = 10^{-4}$ and the maximum absolute data value for each variable. For FPZIP, we set the number of retained bits to 21 as approximate $eb_{rel} = 10^{-4}$ for all the variables. We will use the same test method in the following experiments. 

\begin{table}[]
\centering
\caption{Descriptions of N-body simulation data sets used in the assessment}
\label{tab: apps}
\begin{adjustbox}{width=0.9\columnwidth}
\begin{tabular}{|c|c|c|c|}
\hline
\textbf{Name} & \textbf{\# of Particles} & \textbf{\# of Snapshots} & \textbf{Data Size} \\ \hline
HACC          & 147.3 million            & 500                      & 1.8 TB             \\ \hline
AMDF          & 2.8 million              &  500                        &  34 GB                  \\ \hline
\end{tabular}
\end{adjustbox}
\end{table}

\begin{table}[]
\centering
\caption{Compression ratios of state-of-the-art lossless and lossy compressors on N-body simulation data sets under $eb_{rel} = 10^{-4}$}
\label{tab: overallcr}
\begin{adjustbox}{width=0.5\columnwidth}
\begin{tabular}{|l|c|c|}
\hline
\textbf{Compressor}        & \textbf{HACC} & \textbf{AMDF} \\ \hline
GZIP    & 1.2  &  1.1  \\ \hline
CPC2000 & 3.5  &  \textbf{3.2}  \\ \hline
FPZIP   & 3.1  &  1.8  \\ \hline
ISABELA & 1.4  &  1.2  \\ \hline
ZFP     & 2.3  &  1.9  \\ \hline
SZ      & \textbf{4.6}  & 2.7  \\ \hline
\end{tabular}
\end{adjustbox}
\vspace{-4mm}
\end{table}

Table \ref{tab: overallcr} shows the compression ratios of the current state-of-the-art lossless and lossy compressors on the HACC and AMDF data sets. The table illustrates that for the HACC data sets SZ outperforms the other current state-of-the-art compression methods in regards to the compression ratio; for the AMDF data sets, CPC2000 has the highest compression ratio. These results indicate that different lossy compressors may have  different performance on various N-body simulation data sets, a fact that motivates us to conduct an in-depth research on the best-fit lossy compression methods for N-body simulations. In the following discussion, we exclude GZIP and ISABELA because of their fairly low ratios.

\section{Exploration of Optimizations for Lossy compression on N-body Simulation Data Sets}
\label{sec: optimization}

In this section, we first discuss prediction models for lossy compression on N-body simulation data sets and optimize the prediction model for SZ to improve its compression ratios on N-body simulation data sets. We next propose several techniques for optimizing the lossy compression, including SZ and CPC2000, on the molecular dynamics simulation data sets.
We then explore the best-fit lossy compression methods for cosmology simulation data sets, which are more complicated, such as HACC.

\subsection{Prediction Models of Lossy Compression on N-Body Simulation Data Sets}
A good prediction model is  important for achieving high prediction accuracy and compression ratio for the prediction-based lossy compressors. 

Currently, SZ lossy compressor adopts a multilayer prediction model for multidimensional mesh data sets, and the multilayer prediction model will become the linear-curve-fitting  (LCF) model when the dimensional size is only 1. Specifically, the LCF model uses the previous two data values to predict the data point being compressed. For example, the LCF model predicts the value of $vx_{i}$ by calculating $vx_{i}^{pred} = 2vx_{i-1} - vx_{i-2}$, where $vx_{i-1}$ and $vx_{i-2}$ are the values of two previous points. However, the LCF model may not be applicable to N-body simulation data because of their high irregularity, such as the velocity variables $vx, vy, vz$.

The Lerenzo predictor used in FPZIP degrades into a prediction model, namely, a  last-value (LV) model, when it compresses 1D data sets. Specifically, the LV model uses the same value of the previous one data point to make a prediction, for example, $vx_{i}^{pred} = vx_{i-1}$. The LV model is simple but practical. We calculate the NRMSE of the different variables in the N-body simulation data sets including the HACC and AMDF data sets. Table \ref{tab: predmodel} illustrates that the prediction accuracy of the LV model is higher than the prediction accuracy of the LCF model on both data sets.

\begin{figure}[t]
\centering
\includegraphics[scale=0.34]{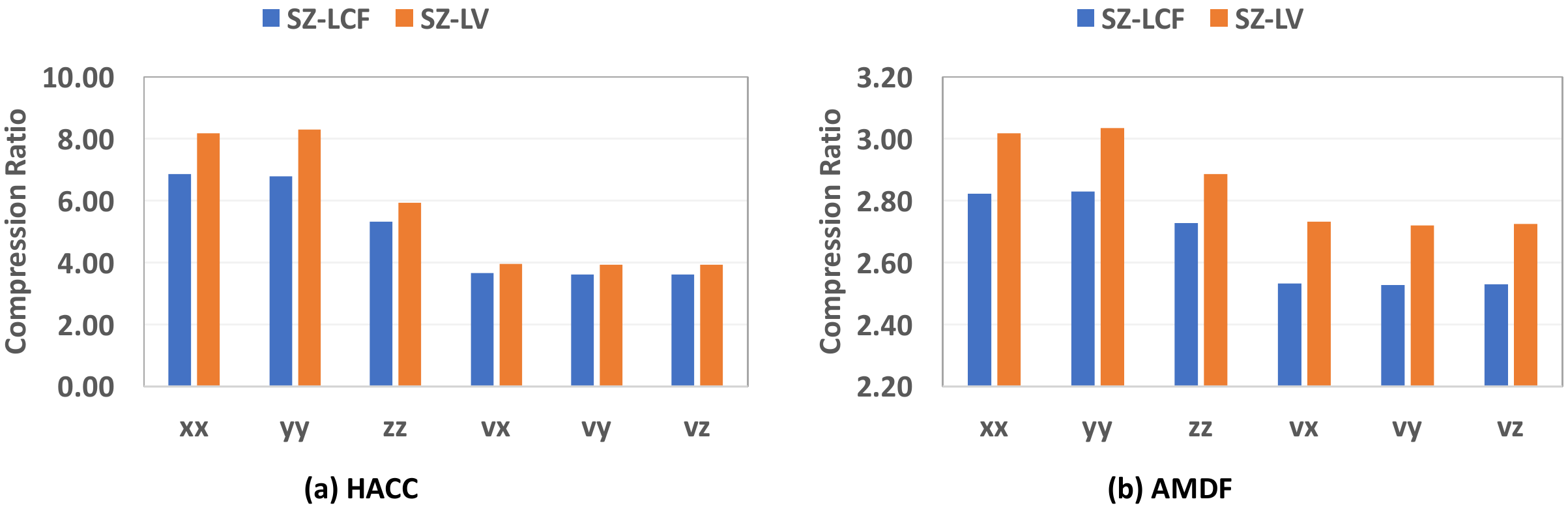}
\caption{Compression ratios of lossy compressor SZ with LCF and LV model on (a) HACC and (b) AMDF data sets under $eb_{rel} = 10^{-4}$.}
\label{fig:fig4}
\end{figure}

Inspired by LV's higher prediction accuracy, we modify the prediction model of SZ from LCF to LV. Figure \ref{fig:fig4} shows the compression ratios of the original SZ (i.e., SZ-LCF) and the improved SZ (i.e., SZ-LV) on the HACC and AMDF data sets. It illustrates that the SZ-LV has higher compression ratios than SZ-LCF on all the variables and the compression ratios are increased by 10.1\% on the average. We note that FPZIP also uses the LV prediction method for 1D data sets but it has lower compression quality than both SZ-LCF and SZ-LV (as shown in Table \ref{tab: overallcr} and Figure \ref{fig:fig4}), because of largely different designs after the prediction. Specifically, FPZIP converts every floating-point value to a binary integer and arithmetically encodes only the leading-zero part of residuals (predicted-values minus real-values). However, the leading-zero part takes a small portion in each residual because the Lorenzo predictor is not accurate on N-body data. The remainder raw bits are not compressed. SZ/SZ-LV adopt linear-scaling quantization approach with a very large number of quantization intervals, such that entropy-coding can be applied to most data of the dataset (e.g. 99\%). We also evaluate the single-core compression rate of SZ-LV based on the execution time and the data size. SZ/SZ-LV have similar/comparable compression rates.

\begin{table}[]
\centering
\caption{Normalized root mean square error of different variables using LCF and LV prediction model on PIC simulation data sets}
\label{tab: predmodel}
\begin{adjustbox}{width=0.5\columnwidth}
\begin{tabular}{|l|c|c|c|c|}
\hline
\multirow{2}{*}{} & \multicolumn{2}{c|}{\textbf{HACC}} & \multicolumn{2}{c|}{\textbf{AMDF}} \\ \cline{2-5} 
                  & \textbf{LCF}     & \textbf{LV}     & \textbf{LCF}      & \textbf{LV}      \\ \hline
\textbf{xx}       & 0.001            & 0.0007          & 0.10              & 0.07             \\ \hline
\textbf{yy}       & 0.003            & 0.002           & 0.10              & 0.06             \\ \hline
\textbf{zz}       & 0.061            & 0.043           & 0.14              & 0.09             \\ \hline
\textbf{vx}       & 0.030            & 0.018           & 0.24              & 0.14             \\ \hline
\textbf{vy}       & 0.032            & 0.020           & 0.25              & 0.14             \\ \hline
\textbf{vz}       & 0.031            & 0.019           & 0.24              & 0.14             \\ \hline
\end{tabular}
\end{adjustbox}
\end{table}

\subsection{Optimizations of Lossy Compression on Molecular Dynamics Simulation Data Sets}
\label{subsec: opt-md}
Sorting is a classic method to enhance the data continuity and has been used in many existing compressors, such as ISABELA \cite{isabela} and CPC2000 \cite{cpc2000}. It suffers from several limitations, however, especially for the snapshot with extremely large numbers of data points. First, sorting is a time-consuming operation even when using the fastest algorithm (quick-sort). Second, because of the reordered location of the data in the sorted array, the compressor must use an extra index array to record the original location for each point, which significantly limits the compression ratio. However, storing the extra index information is not necessary for particle elements, since, unlike regular multidimensional mesh data, the particle elements in each array are allowed to be reordered in the reconstructed data set as long as the locations or indices of the elements with regard to the same particle can be consistent across arrays. We usually perform the sorting once on one data array and adjust the indices with the same orders on other data arrays. CPC2000 proposes a useful technique, namely R-index based sorting, which matches the above conditions.

\begin{figure}[t]
\centering
\includegraphics[scale=0.31]{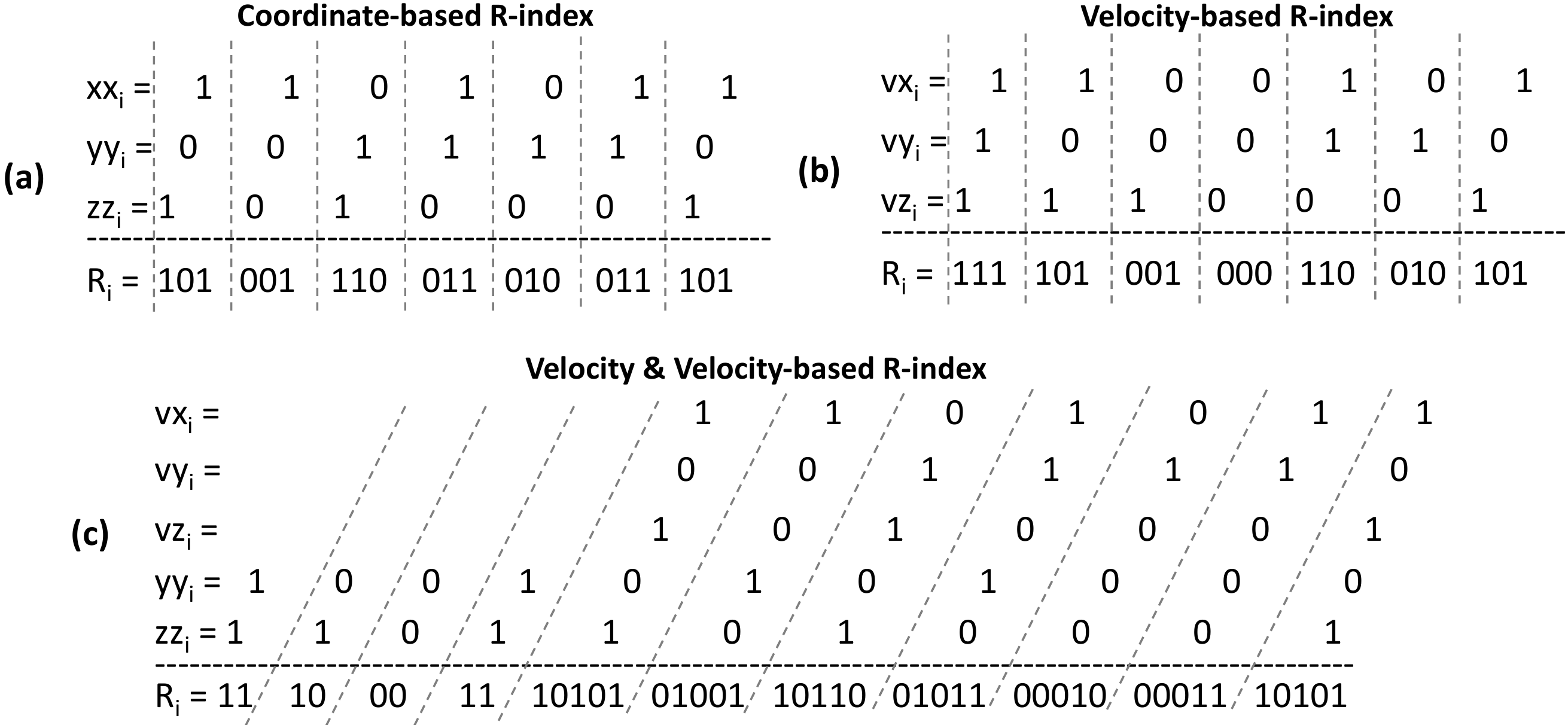}
\caption{Construction of R-index based on (a) coordinate variables and (b) coordinate + velocity variables.}
\label{fig:fig3}
\vspace{-2mm}
\end{figure}

\begin{figure}[t]
\centering
\includegraphics[scale=0.40]{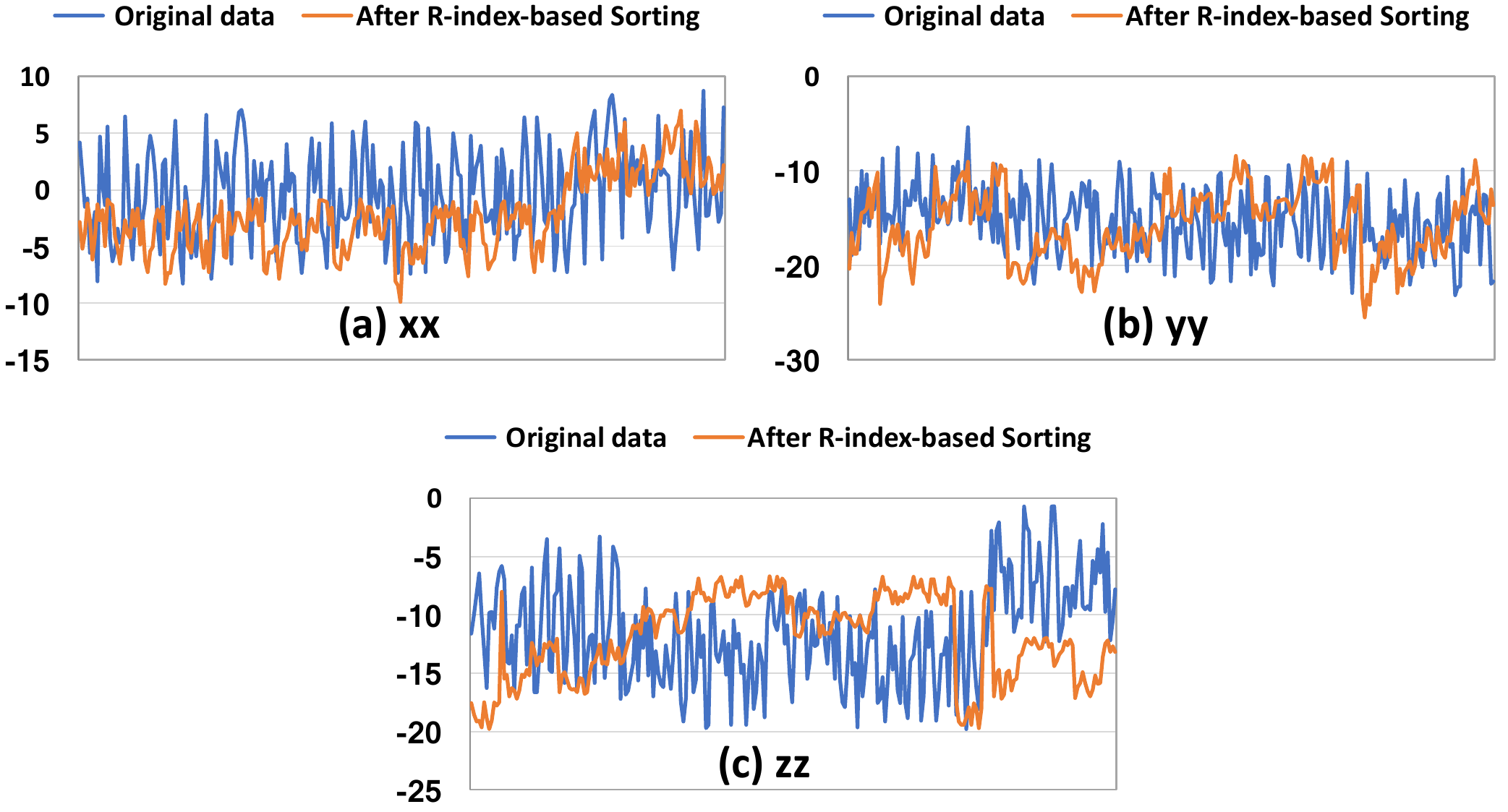}
\caption{Partial visualization of coordinate variables before and after R-index-based sorting.}
\label{fig:fig5}
\end{figure}

CPC2000 first converts coordinate variables from floating-point values to integer numbers by dividing them by a user-set error bound. Then it generates an R-index (i.e., $R_{i}$) by interleaving the binary representations of three coordinate variables ($xx_{i}, yy_{i}, zz_{i}$), as illustrated in Figure \ref{fig:fig3} (a). After that, CPC2000 sorts all the variables, including coordinate and velocity variables, based on the R-index values constructed from coordinates. Figure \ref{fig:fig5} shows a portion of coordinate variables before and after R-index-based sorting. It demonstrates that although R-index-based sorting cannot completely sort every value in each coordinate variable, as can traditional sorting (e.g., quick-sort), the converted data points will become more ordered and thus smoother. 
Even though CPC2000 does not necessarily store the extra index information, it still suffers from low compression rate. For example, the single-core compression rate of SZ-LV for compressing AMDF data is about 94.4 MB/s, whereas the CPC2000's is only about 21.5 MB/s.

Thus, we propose an optimization strategy, namely SZ-LV-PRX. It can achieve the same compression ratio as CPC2000's but significantly improve the compression rate as twice as much. The method comprises two steps: (1) performing a partial-radix sorting to sort the R-index array such that the compression performance can be improved significantly, and then (2) applying SZ-LV lossy compression method on the reordered data arrays ($xx$, $yy$, $zz$) instead of compressing R-index directly as done by CPC2000.

\begin{table}[]
\centering
\caption{Compression ratios and rates using SZ-LV and R-index-based sorting technique with different segment sizes on AMDF data sets under $eb_{rel} = 10^{-4}$}
\label{tab: SZ-LV-RX}
\begin{adjustbox}{width=0.72\columnwidth}
\begin{tabular}{|l|c|c|c|}
\hline
                                     & \textbf{\begin{tabular}[c]{@{}c@{}}Segment\\ Size\end{tabular}} & \textbf{\begin{tabular}[c]{@{}c@{}}Compression\\ Ratio\end{tabular}} & \textbf{\begin{tabular}[c]{@{}c@{}}Compression\\ Rate (MB/s)\end{tabular}} \\ \hline
\textbf{SZ-LV}                       & /                                                               & 2.85                                                                 & 94.4                                                           \\ \hline
\multirow{5}{*}{\textbf{SZ-LV-RX}} & 1024                                                            & 3.03                                                                 & 36.0                                                          \\ \cline{2-4} 
                                     & 2048                                                            & 3.07                                                                 & 33.9                                                          \\ \cline{2-4} 
                                     & 4096                                                            & 3.11                                                                 & 34.3                                                          \\ \cline{2-4} 
                                     & 8192                                                            & 3.15                                                                 & 33.6                                                          \\ \cline{2-4} 
                                     & 16384                                                           & 3.20                                                                 & 35.0                                                          \\ \hline
\end{tabular}
\end{adjustbox}
\vspace{-3mm}
\end{table}

In what follows, we describe step 2 first, as the design motivation of step 1 is based on step 2. Specifically, the objective of step 1 is just to mitigate the execution overhead of step 2.

In the step 2, we split the data set into multiple segments, construct the R-index array based on $xx$, $yy$ and $zz$ in each segment, and then perform a radix sorting on the R-index array by segment. Table \ref{tab: SZ-LV-RX} shows the compression ratios and the compression rates when compressing the AMDF's data arrays reordered by R-index sorting method with various segment sizes, under a value-range-based relative error bound $10^{-4}$. The table demonstrates that we can increase the compression ratio for SZ-LV to 3.2 by using the segment size of 16,384, but the compression rate is decreased to 46\% (from 94.4 MB/s to 35.0 MB/s). Note that in the table ``RX'' represent R-indeX-based sorting.

\begin{table}[]
\centering
\vspace{-2mm}
\caption{Compression ratios and rates using SZ-LV-PRX technique with different ignored bits on AMDF data sets under $eb_{rel} = 10^{-4}$}
\label{tab: SZ-LV-PRX}
\begin{adjustbox}{width=.85\columnwidth}
\begin{tabular}{|l|c|c|c|c|}
\hline
                                                                                   & \textbf{\begin{tabular}[c]{@{}c@{}}Segment\\ Size\end{tabular}} & \textbf{\begin{tabular}[c]{@{}c@{}}Ignored \\ Bits\end{tabular}} & \textbf{\begin{tabular}[c]{@{}c@{}}Compression\\ Ratio\end{tabular}} & \textbf{\begin{tabular}[c]{@{}c@{}}Compression\\ Rate (MB/s)\end{tabular}} \\ \hline
\textbf{SZ-LV}                                                                     & /                                                               & /                                                               & 2.85                                                                 & 94.4                                                           \\ \hline
\multirow{5}{*}{\textbf{\begin{tabular}[c]{@{}l@{}}SZ-LV-PRX\end{tabular}}} & 16384                                                           & 0                                                               & 3.20                                                                 & 35.0                                                          \\ \cline{2-5} 
                                                                                   & 16384                                                           & 2                                                               & 3.20                                                                 & 36.6                                                          \\ \cline{2-5} 
                                                                                   & 16348                                                           & 4                                                               & 3.20                                                                 & 39.4                                                          \\ \cline{2-5} 
                                                                                   & \textbf{16384}                                                  & \textbf{6}                                                      & \textbf{3.20}                                                        & \textbf{43.8}                                                 \\ \cline{2-5} 
                                                                                   & 16384                                                           & 8                                                               & 3.16                                                                 & 51.1                                                          \\ \hline
\end{tabular}
\end{adjustbox}
\vspace{-4mm}
\end{table}

As for the first step, we improve the compression rate by performing a partial-radix sorting. Specifically, we ignore the last few least significant digits and start the radix sorting from the last n-th bit. Note that for the R-index constructed from the three coordinate variables, we sort it by three bits at each round. For example, in Figure \ref{fig:fig4} (a), we start the radix sorting from the last third 3-bit, which is ``010'';  this process can be described as Partial R-indeX (PRX) based sorting by ignoring the last two 3-bits. 

The PRX sorting will keep a high smoothness of the reordered arrays for compression while suffering from low time. We evaluate the SZ-LV compression based on the PRX sorting on the AMDF data sets. Table \ref{tab: SZ-LV-RX} shows the compression ratios and rates with different numbers of ignored bits using the same segment size 16,384 under the value-range-based relative error bound $10^{-4}$. The \emph{ignored bits} in Table \ref{tab: SZ-LV-PRX} represent the number of ending 3-bits overlooked in the R-index. For example, the number of ignored bits being set to 2 represents that we ignore the right-most two 3-bits in the R-index before conducting the radix sorting. The table shows that our proposed optimization method can significantly improve the compression rate from 35.0 MB/s to 43.8 MB/s while maintaining the compression ratio of 3.20. The reason the partial radix sorting may not degrade the compression ratio is that the reordered data arrays actually still exhibit local irregularity within a relatively small index range even in the full-radix R-index sorting method. That is, the last few bits in R-index do not need to be sorted, in order to keep the smoothness of the reordered arrays. 

\begin{figure}[t]
\centering
\includegraphics[scale=0.46]{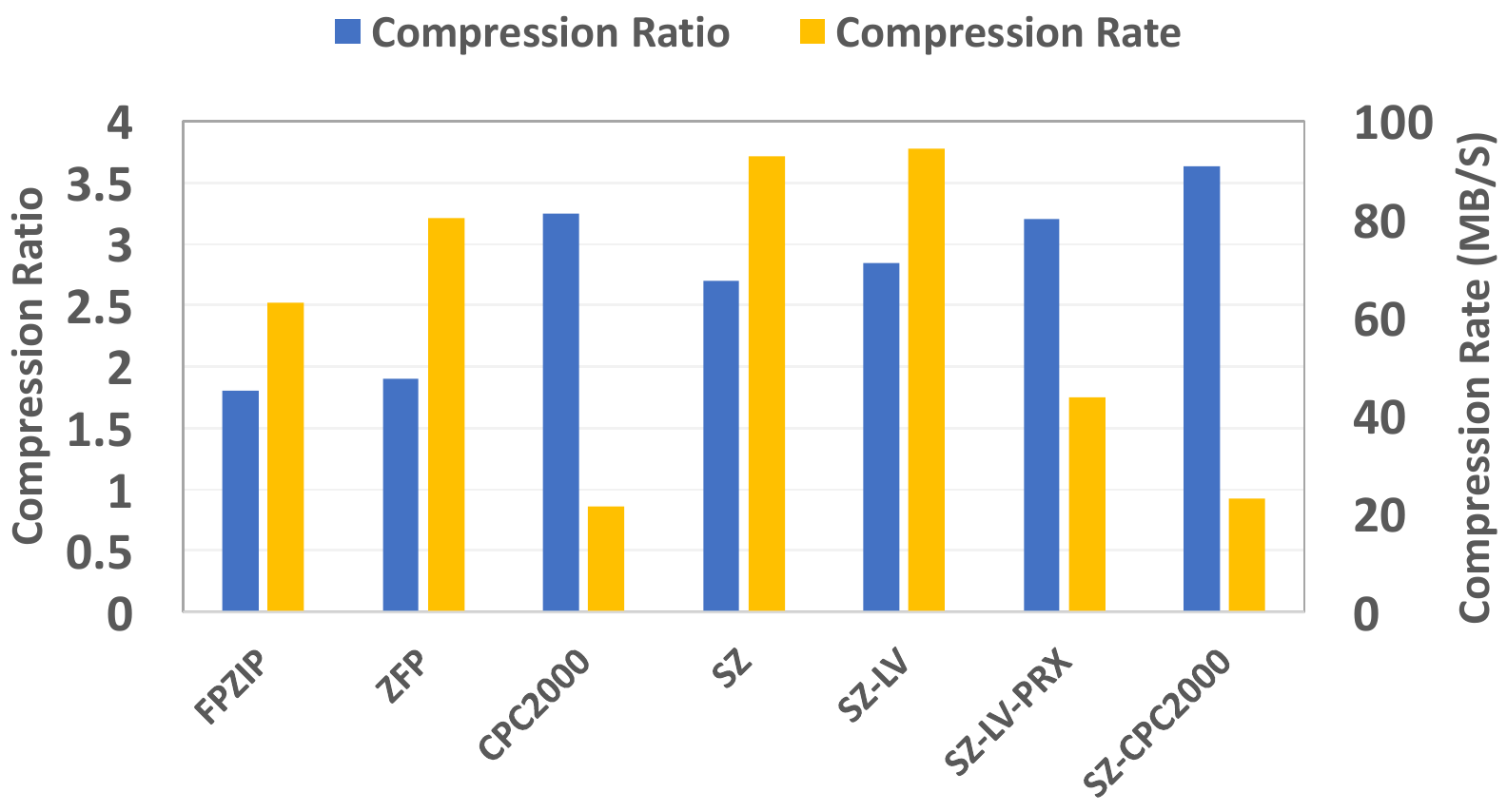}
\caption{Comparison of compression ratios and rates with different lossy compression methods on AMDF data sets under $eb_{rel} = 10^{-4}$.}
\label{fig:fig2}
\end{figure}

Based on our observation, CPC2000's compression ratio is 2x higher than SZ's on the coordinate variables, so we propose another optimization method, namely SZ-CPC2000, to further improve the compression ratio on molecular dynamics simulation data sets.
After converting the floating-point data of velocity variables to the integer numbers, CPC2000 compresses the integer velocity values by applying an adaptive variable-length coding method \cite{cpc2000}, which adopts status bits to differentiate the different values in the bit stream. However, such a variable-length encoding method suffers from relatively high status bit overhead (1$\sim$10 bits per value based on our experimental observations), leading to a very limited compression ratio. So we propose SZ-CPC2000 combining SZ-LV and CPC2000. It applies SZ-LV technique with a tailored Huffman encoding \cite{sz17} on the compression of velocity data that was reordered by R-index sorting. Figure \ref{fig:fig2} demonstrates that CPC2000 has the highest compression ratio from among all existing lossy compressors (including FPZIP, ZFP, SZ) on the molecular dynamics simulation data. Our designed SZ-CPC2000 can further improve the compression ratio by about 13\% and improve the compression rate by about 10\%. Also note that our proposed method SZ-LV obtains the best compression rate among all the lossy compression methods.

\subsection{Exploration of Lossy Compression on Cosmology Simulation Data Sets}
\label{subsec: opt-hacc}

In this subsection, we continue exploring an optimized lossy compression method on the cosmology simulation data sets. Table \ref{tab: overallcr} shows that SZ and CPC2000 are the two most promising lossy compressors on the HACC data sets. In particular, our proposed SZ-LV outperforms CPC2000 in each variable of the HACC data as shown in Table \ref{tab: velRX}. Thus, we use SZ-LV as the baseline compression model and try to further optimize its compression quality for the HACC data. 

\begin{table}[]
\centering
\caption{Compression Ratios of several R-index attempts on HACC data}
\label{tab: velRX}
\begin{adjustbox}{width=\columnwidth}
\begin{tabular}{|l|c|c|c|c|c|}
\hline
                 & \textbf{CPC2000} & \textbf{SZ-LV} & \textbf{\begin{tabular}[c]{@{}c@{}}SZ-LV + \\ Coordinate-based\\ R-index\end{tabular}} & \textbf{\begin{tabular}[c]{@{}c@{}}SZ-LV + \\ Velocity-based\\ R-index\end{tabular}} & \textbf{\begin{tabular}[c]{@{}c@{}}SZ-LV + \\ Coordinate \& Velocity-\\ based index\end{tabular}} \\ \hline
\textbf{xx}      & 7.1              & 8.18           & 8.55                                                                                   & 8.00                                                                                   & 8.14                                                                                              \\ \hline
\textbf{yy}      & 7.1              & 8.31           & 6.28                                                                                   & 5.07                                                                                   & 5.62                                                                                              \\ \hline
\textbf{zz}      & 7.1              & 5.93           & 6.73                                                                                   & 3.66                                                                                   & 5.15                                                                                              \\ \hline
\textbf{vx}      & 2.3              & 3.97           & 3.84                                                                                   & 4.36                                                                                   & 4.21                                                                                              \\ \hline
\textbf{vy}      & 2.3              & 3.92           & 3.79                                                                                   & 4.48                                                                                   & 4.25                                                                                              \\ \hline
\textbf{vz}      & 2.3              & 3.93           & 3.87                                                                                   & 4.69                                                                                   & 4.38                                                                                              \\ \hline
\textbf{Overall} & 3.5              & 5.12           & 4.97                                                                                   & 4.76                                                                                   & 5.02                                                                                              \\ \hline
\end{tabular}
\end{adjustbox}
\vspace{-3mm}
\end{table}

Similar to the first optimization method proposed in Section \ref{subsec: opt-md}, two candidate solutions are reorganizing the data in terms of velocity-based R-index sorting and a coordinate\&velocity-based R-index sorting, as shown in Figure \ref{fig:fig3} (b) and (c), respectively. The construction is similar to the traditional one as shown in Figure \ref{fig:fig3} (a). Table \ref{tab: velRX} presents the compression ratios of SZ-LV plus these three types of R-index with the segment size of 4096. It demonstrates that all three R-index-based sorting strategies are unable to improve the compression ratio for the HACC data sets. The table shows that SZ-LV plus velocity-based R-index can improve the compression ratios of velocity variables about 20\% on average, but the compression ratios of coordinate variables drop significantly, especially $yy$ (from 8.3 to 5.1) and $zz$ (from 5.9 to 3.7), and counteract the improvement from velocity variables. After we put both the coordinate and velocity variables into the construction of the R-index, the compression ratio of $zz$ is improved by more than 40\% (only 13\% lower than using the SZ-LV without R-index-based sorting), whereas $yy$ is still much lower than using SZ-LV without R-index-based sorting (i.e., 32\%). Therefore, the overall compression ratio using coordinate\&velocity-based index is still worse than without R-index-based sorting. We observe that the compression ratio of $xx$, $zz$, $vx$, $vy$, $vz$ can be improved at least by one type of R-index sorting, but the compression ratio of $yy$ is always reduced. Therefore, for HACC that performs N-body simulations with a hierarchical model, SZ-LV plus R-index sorting fail to improve the compression ratio of the whole data sets. This is due to the fact that unlike AMDF, not all variables in HACC are very disordered: in particular, $yy$ is actually approximately sorted in an increasing order in a wide-index range, such that any attempt of reordering the variables, such as R-index based sorting, will get it disordered unexpectedly, leading to the low compression ratios on them in turn. Therefore, the orderly variable with high autocorrelation is not applicable to be reordered by any R-index based sorting methods, such as SZ-LV-PRX and SZ-CPC2000. Consequently, our improved SZ-LV is the best-ratio lossy compression method in this case. In the next section, we will show that SZ-LV is the best lossy compressor with respect to both compression ratio and rate for cosmology simulation based on the HACC data sets.

\section{Empirical Evaluation}
\label{sec: evaluation}

In this section, we evaluate our proposed optimization solutions on the HACC and AMDF simulation data sets (as shown in Table \ref{tab: apps}) using 1024 cores (i.e., 64 nodes, each with two Intel Xeon E5-2670 processors and 64 GB DDR3 memory, and each processor with 16 cores) from the Blues supercomputer at Argonne and compare with the state-of-the-art lossy compressors. The details of the data sets are described in Table \ref{tab: apps}. 
The storage system of Blues uses General Parallel File Systems (GPFS). The different directories on the cluster are located on the separate GPFS file systems that are shared by all nodes on the cluster. These file systems are located on a raid array and are served by multiple file servers. The I/O and storage systems are typical of high-end supercomputer facilities. We use the same test methods described in Section \ref{sec: assess} for the state-of-the-art lossy compressors including CPC2000, FPZIP, ZFP, and SZ. 

We first perform the evaluation on the full 1.8 TB HACC data sets in parallel. As we analyzed in Section \ref{subsec: opt-hacc}, R-index-based compression methods, such as CPC2000, SZ-LV-PRX, and SZ-CPC2000, are not applicable for the HACC data sets, therefore, we focus on evaluating FPZIP, ZFP, and SZ-LV on the HACC data sets in parallel. In the experiments, we increase the scale from 1 node to 64 nodes with 16 processes on each node in the cluster. Each process will independently compress a single snapshot of HACC data within its memory. We note that the decompression rates of FPZIP, ZFP, and SZ are not lower than their compression rates because of less steps in decompression, as demonstrated in \cite{fpzip, zfp, sz17}. Specifically, FPZIP's decompression rate is similar to its compression rate \cite{fpzip}; ZFP's decompression rate is 1.4x of its compression rate \cite{zfp}; and SZ's decompression rate is 1.7x of its compression rate \cite{sz17}. We confirm it in the experiments. We also confirm that SZ-LV and SZ have the similar compression and decompression rate. So we only present the compression rate due to space limitations. 

\begin{figure}
\centering
   \begin{subfigure}[b]{0.40\textwidth}
   \includegraphics[width=\textwidth]{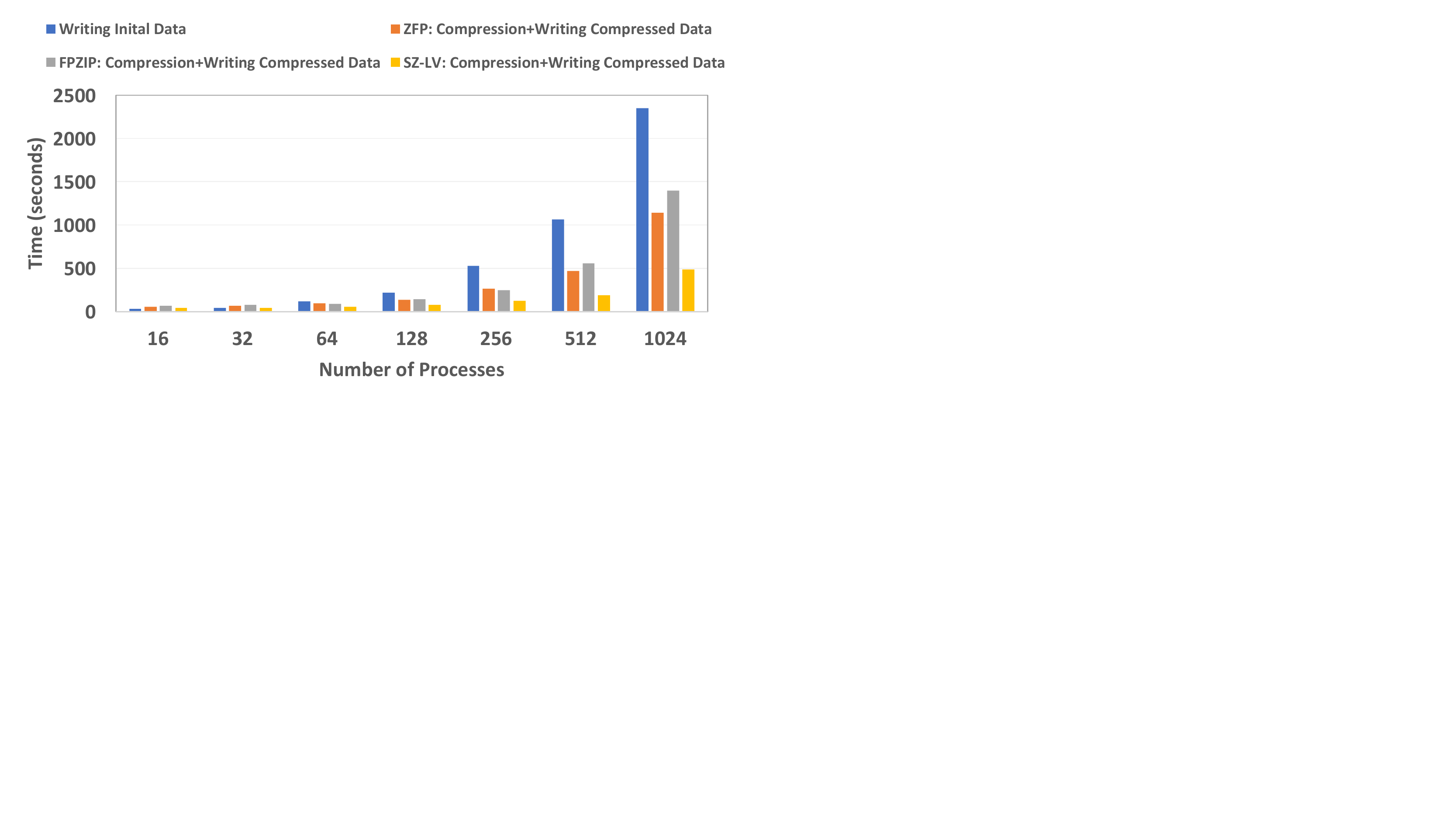}
   \caption{}
   \label{fig:fig7a} 
\end{subfigure}

\begin{subfigure}[b]{0.40\textwidth}
   \includegraphics[width=\textwidth]{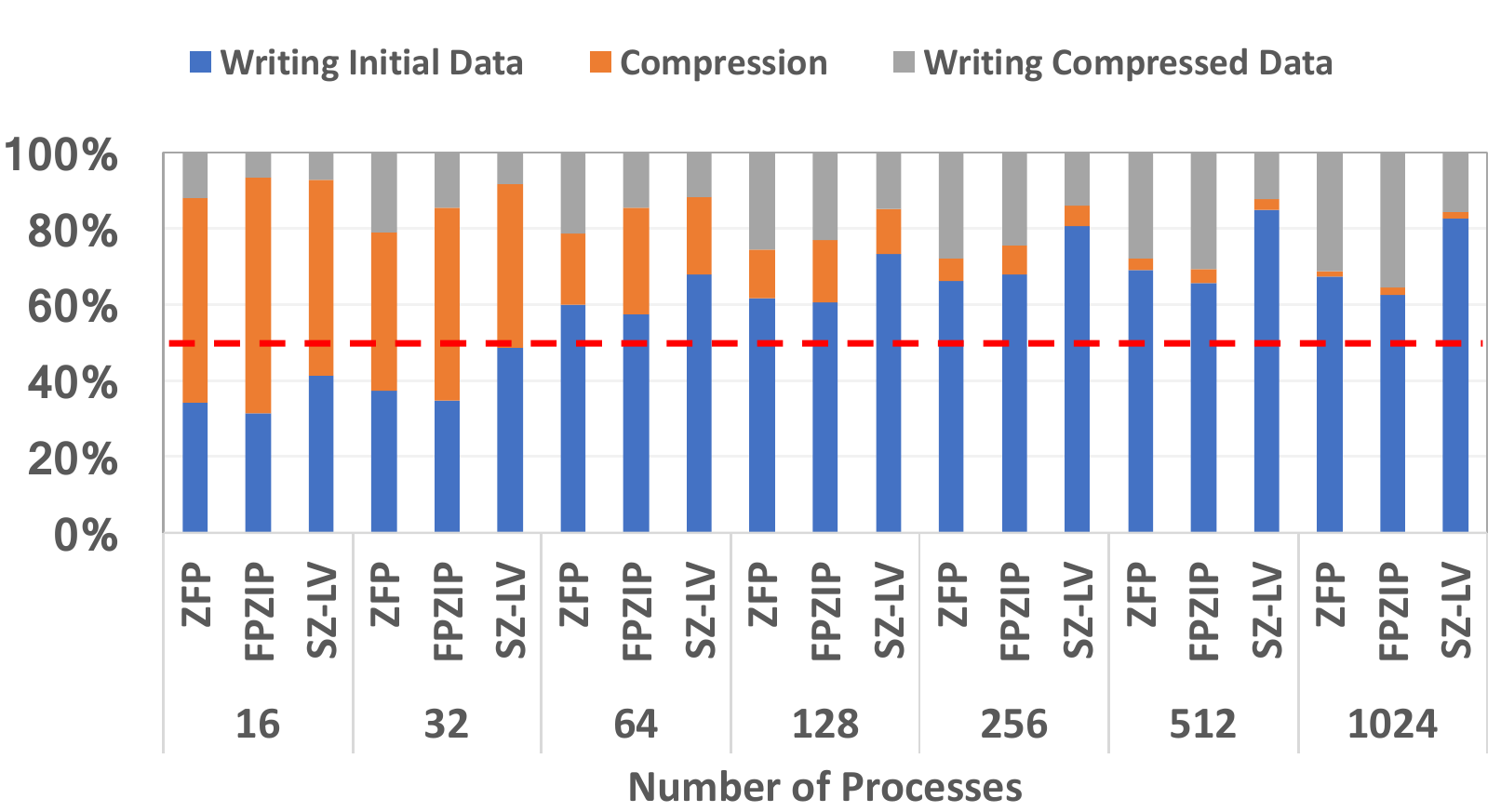}
   \caption{}
   \label{fig:fig7b}
\end{subfigure}

\caption{Comparison of time to compress and write compressed data against time to write original data with ZFP, FPZIP, and SZ-LV on HACC data sets.}
\label{fig:fig7}
\vspace{-1mm}
\end{figure}

Figure \ref{fig:fig7} presents the comparison of the time to write initial data and the times to write compressed data plus compression with ZFP, FPZIP, and SZ-LV respectively. It shows that the time to write initial data is always much longer than the time to write compressed bytes plus the time of compression when the number of processes is 64 or more, no matter which compressor is used. Hence, the parallel applications can save I/O time from 64 processes on this system by using the lossy compression in this parallel environment. Figure \ref{fig:fig7} (a) shows that SZ-LV always outperforms the other two lossy compressors because of its higher compression ratio and rate. In Figure \ref{fig:fig7} (b), we further present the compression time and the time to write compressed data. Each bar includes time of compression, time to write compressed data, and the time to write the initial data. We normalize the sum to 100\% and plot a dash line at 50\% to ease the comparison. It shows that SZ-LV can reduce I/O time by 80\% and outperforms the second-best solution by 60\% with 1024 cores. Note that the relative time spent in I/O will keep increasing with the number of processes due to inevitable bottleneck of the I/O bandwidth when writing data simultaneously by many processes. But the parallel in situ compression has a nearly linear speedup with the number of processes, which indicates the performance gains will be greater with increasing scale.

We present in Table \ref{tab: throughput} the scalability and compression rates (GB/s) of the three lossy compression methods without I/O in different scales ranging from 1 to 1024 processes on the cluster. We measure the time of compression without I/O time and use the maximum time among all the processes. We can see that the parallel efficiency of these three lossy compression methods can always stay nearly 100\% from 1 to 256 processes, which demonstrates that the in situ compression with FPZIP, ZFP, and SZ-LV have linear speedup with the number of processes. However, the parallel efficiency is decreased to about 85\% when the total number of processes is greater than 256. This performance degradation is due to node internal limitations when multiple cores share the memory on each node.

\begin{table}[]
\centering
\caption{Compression rate (GB/s) and parallel efficiency of ZFP, FPZIP, and SZ-LV on HACC data sets.}
\label{tab: throughput}
\begin{adjustbox}{width=\columnwidth}
\begin{tabular}{|l|c|c|c|c|c|c|}
\hline
\multirow{2}{*}{\textbf{\begin{tabular}[c]{@{}l@{}}Number of\\ Processes\end{tabular}}} & \multicolumn{2}{c|}{\textbf{ZFP}}                                                                                           & \multicolumn{2}{c|}{\textbf{FPZIP}}                                                                                         & \multicolumn{2}{c|}{\textbf{SZ-LV}}                                                                                            \\ \cline{2-7} 
                                                                                        & \begin{tabular}[c]{@{}c@{}}Comp\\ Rate\end{tabular} & \begin{tabular}[c]{@{}c@{}}Parallel\\ Efficiency\end{tabular} & \begin{tabular}[c]{@{}c@{}}Comp\\ Rate\end{tabular} & \begin{tabular}[c]{@{}c@{}}Parallel\\ Efficiency\end{tabular} & \begin{tabular}[c]{@{}c@{}}Comp\\ Rate\end{tabular} & \begin{tabular}[c]{@{}c@{}}Parallel\\ Efficiency\end{tabular} \\ \hline
\textbf{1}                                                                              & 0.17                                                        & /                                                             & 0.13                                                        & /                                                             & 0.22                                                        & /                                                             \\ \hline
\textbf{16}                                                                             & 2.72                                                        & 100\%                                                         & 2.16                                                        & 100\%                                                         & 3.44                                                         & 99.5\%                                                        \\ \hline
\textbf{32}                                                                             & 5.42                                                        & 99.6\%                                                        & 4.20                                                        & 99.4\%                                                        & 6.88                                                         & 99.5\%                                                        \\ \hline
\textbf{64}                                                                             & 10.8                                                       & 98.8\%                                                        & 8.38                                                        & 99.1\%                                                        & 13.7                                                        & 99.1\%                                                        \\ \hline
\textbf{128}                                                                            & 21.6                                                       & 99.3\%                                                        & 16.80                                                       & 99.4\%                                                        & 27.3                                                        & 98.7\%                                                        \\ \hline
\textbf{256}                                                                            & 43.2                                                       & 99.3\%                                                        & 33.4                                                       & 98.9\%                                                        & 54.7                                                        & 99.0\%                                                        \\ \hline
\textbf{512}                                                                            &    80.0                                                       &    91.9\%                                                          &        67.1                                                    &    99.3\%                                                           &      111.1                                                     &   99.1\%                                                             \\ \hline
\textbf{1024}                                                                           &  146.2                                                           &    83.8\%                                                           &     119.2                                                        &    88.2\%                                                           &    194.6                                                         &   88.0\%                                                            \\ \hline
\end{tabular}
\end{adjustbox}
\vspace{-5mm}
\end{table}

We then evaluate the lossy compression on the AMDF data sets, including the state-of-the-art lossy compressors (i.e., CPC2000, FPZIP, ZFP, SZ) and our proposed optimization solutions (i.e., SZ-LV, SZ-LV-PRX, SZ-CPC2000). We evaluate the compression rate on a single core 
of the cluster as shown in Figure \ref{fig:fig2}. It shows that (1) our proposed method SZ-LV can lead to the best rate with only 12\% lower compression ratio than the best-ratio compressor CPC2000, (2) our proposed method SZ-LV-PRX can obtain 2x improvement on the compression rate with unchanged compression ratio compared with CPC2000, and (3) our proposed method SZ-CPC2000 can further improve the compression ratio by 13\% with 10\% improvement on the compression rate over CPC2000. Therefore, similar to GZIP \cite{gzip}, we can set three modes---\texttt{best\_speed}, \texttt{best\_tradeoff}, and \texttt{best\_compression} -- for the lossy compression on the molecular dynamics simulation data based on our proposed optimization methods. Specifically, we can set SZ-LV as the \texttt{best\_speed} mode, SZ-LV-PRX as the \texttt{best\_tradeoff} mode, and  SZ-CPC2000 as the \texttt{best\_compression}  mode.

We next evaluate the maximum compression errors for all the lossy compressors in our experiments, including FPZIP, ZFP, CPC2000, SZ, SZ-LV, and SZ-CPC2000, on the HACC and AMDF data sets. The maximum compression errors of CPC2000, SZ, SZ-LV, SZ-LV-PRX, and SZ-CPC2000 are exactly equal to the user-set error bound, while ZFP's maximum compression errors are lower than the user-set error bound. For example, ZFP's maximum compression error of each variable ranges from 3.2E-5 to 4.6E-5 when $eb_{rel}$ is set to 1E-4, which means ZFP over-preserves the compression errors in respect of user's accuracy requirement. Since FPZIP is designed for fixed bit-rate, we set the number of retained bits to 21 as approximate value-range-based relative error bound of 1E-4 in the experiments and its maximum compression errors are a bit higher than 1E-4, ranging from 0.6E-4 to 2.4E-4. Users can achieve higher compression accuracy by increasing the number of retained bits in each value.

\begin{figure}
\centering
   \begin{subfigure}[b]{0.39\textwidth}
   \includegraphics[width=\textwidth]{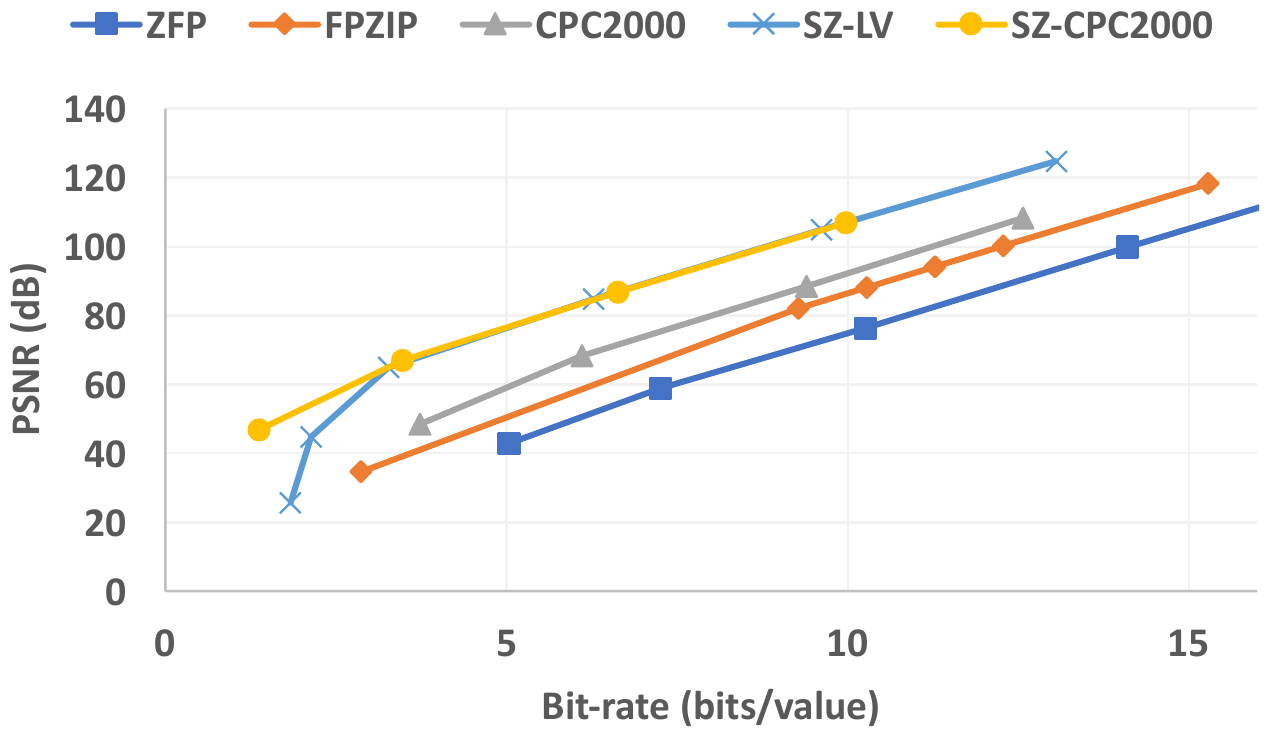}
   \caption{HACC}
   \label{fig:fig9a} 
\end{subfigure}

\begin{subfigure}[b]{0.37\textwidth}
   \includegraphics[width=\textwidth]{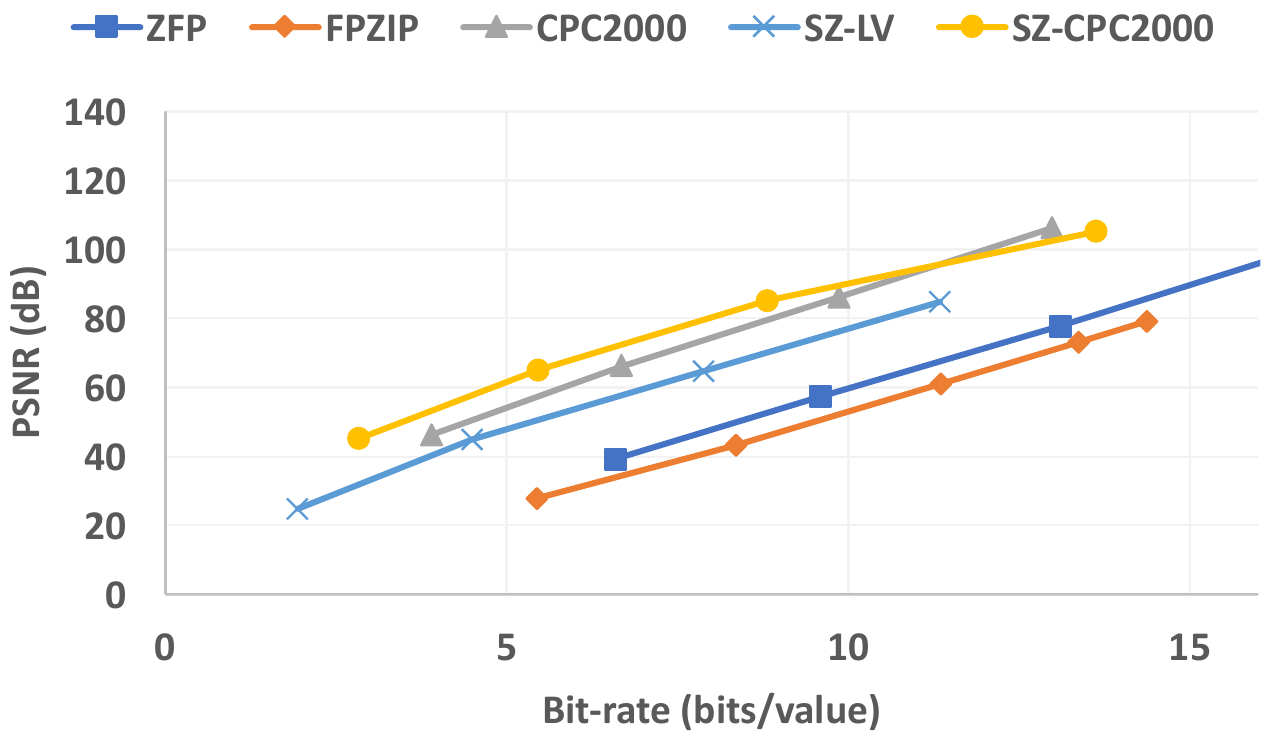}
   \caption{AMDF}
   \label{fig:fig9b}
\end{subfigure}

\caption{Rate-distortion of ZFP, FPZIP, CPC2000, SZ-LV, and SZ-CPC2000 on (a) HACC and (b) AMDF data sets.}
\label{fig:fig9}
\vspace{-6mm}
\end{figure}

At last, we evaluate the distortion quality  for all the lossy compression methods on the HACC and AMDF data sets by using Z-checker \cite{zchecker}. Here rate means bit-rate in bits/value, and we use the peak signal-to-noise ratio (PSNR) to measure the distortion quality. PSNR is calculated as $20 \cdot log_{10}($NRMSE$)$: the higher the better. 
We run ZFP, CPC2000, SZ-LV, and SZ-CPC2000 with different error bounds and run FPZIP with different numbers of retained bits.
Then, we calculate their correponding average bit rates (i.e., 32/compression-ratio) and distortion values (i.e., PSNRs) and plot rate-distortion curves by connecting the scatters. 
Figure \ref{fig:fig9} shows the rate-distortion curves of the different lossy compressors on the two data sets. It illustrates that our proposed method SZ-CPC2000 has the best rate-distortion among all the lossy compression methods on  both the HACC and AMDF data sets when the bit-rate is smaller than 10 bits/value (i.e., compression ratio is higher than 3.2). Note that we test and show the case only with the bit-rate lower than 16 bits/value for the two single-precision data sets, i.e., the compression ratios are higher than 2. As demonstrated in \cite{ratana}, lossless compression can provide a compression ratio up to 2, so it is reasonable to assume that users are interested in lossy compression only if the compression ratio is higher than 2.
\section{Conclusion}
\label{sec: conclusion}

In this work, we explored the lossy compression techniques for single-snapshot data produced by N-body simulations. 
We first assessed state-of-the-art compression techniques on two real-world N-body applications: HACC and AMDF. 
Some of the compression techniques are state-of-the-art tools such as SZ and ZFP; and others (such as CPC2000) are implemented and improved rigorously by us. 
To the best of our knowledge, this is the first attempt to comprehensively assess the quality of lossy compression techniques with a single snapshot for N-body simulations. 
We then designed a series novel optimization techniques.
\begin{itemize}[noitemsep,topsep=4pt]
\item For molecular dynamics simulations, our proposed three compression methods can either lead to the best compression rate or best compression ratio compared with existing state-of-the-art compression techniques.
\item For cosmology simulations, our proposed method SZ-LV is the best lossy compression technique with respect to both compression ratio and rate. Broadly speaking, SZ-LV is more suitable than SZ-LV-PRX/SZ-CPC2000 on the orderly data sets with high autocorrelations.
\item Experiments show that our proposed method SZ-LV can be used as an in situ compression method in a parallel cosmology simulation to reduce the I/O time by more than 80\% with 1024 cores and outperforms the second-best solution by 60\%. Our proposed method SZ-CPC2000 has the best rate-distortion with reasonable compression errors compared with state-of-the-art compression methods.
\end{itemize}
Our optimizations solutions are generic and can be applied to any other N-body scientific data sets. We plan to evaluate on more N-body applications in the future.

\section*{Acknowledgments}
\footnotesize
This research was supported by the Exascale Computing Project (ECP), Project Number: 17-SC-20-SC, a collaborative effort of two DOE organizations - the Office of Science and the National Nuclear Security Administration, responsible for the planning and preparation of a capable exascale ecosystem, including software, applications, hardware, advanced system engineering and early testbed platforms, to support the nation's exascale computing imperative. 

\bibliographystyle{abbrv}
\bibliography{bib/refs}

\end{document}